\documentclass[showkeys, showpacs]{revtex4}
\usepackage{indentfirst}
\begin{document}
\title{\bf Nonstrange baryonia with the open charm}
\author{S.M. Gerasyuta}
\email{gerasyuta@SG6488.spb.edu}
\author{E.E. Matskevich}
\email{matskev@pobox.spbu.ru}
\affiliation{Department of Theoretical Physics, St. Petersburg State University, 198904,
St. Petersburg, Russia}
\affiliation{Department of Physics, LTA, 194021, St. Petersburg, Russia}
\begin{abstract}
The relativistic six-quark amplitudes of the nonstrange baryonia with the
open charm are calculated. The poles of these amplitudes determine the
masses of baryonia. 9 masses of baryonia are predicted.
\end{abstract}
\maketitle
Hadron spectroscopy has always played an important role in the revealing
mechanisms underlying the dynamic of strong interactions.

The heavy hadron containing a single heavy quark is particularly
interesting. The light degrees of freedom (quarks, antiquarks and gluons)
circle around the nearby static heavy quark. Such a system behaves as
the QCD analog of familar hydrogen bound by the electromagnetic
interaction.

In Refs. \cite{1, 2} relativistic generalization of the three-body
Faddeev equations was obtained in the form of dispersion relations in the
pair energy of two interacting quarks. The mass spectrum of $S$-wave
baryons including $u$, $d$, $s$ quarks was calculated by a method based on
isolating the leading singularities in the amplitude. We searched for the
approximate solution of integral three-quark equations by taking into
account two-particle and triangle singularities, all the weaker ones being
neglected. If we considered such an approximation, which corresponds to
taking into account two-body and triangle singularities, and defined all
the smooth functions of the subenergy variables (as compared with the
singular part of the amplitude) in the middle point of the physical region
of Dalitz-plot, then the problem was reduced to the one of solving a system
of simple algebraic equations.

In Ref. \cite{3} the relativistic six-quark equations are found
in the framework of coupled-channel formalism. The dynamical mixing between
the subamplitudes of hexaquark are considered. The six-quark amplitudes
of dibaryons are calculated. The poles of these amplitudes determine the
masses of dibaryons. We calculated the contribution of six-quark
subamplitudes to the hexaquark amplitudes.

In the present paper the six-quark equations for the nonstrange baryonia
with the open charm are found. The nonstrange baryonia $B\bar B_c$ are
constructed without the mixing of the quarks and antiquarks. The
six-quark amplitudes of baryonia are constructed. The relativistic
six-quark equations are obtained in the form of the dispersion relations
over the two-body subenergy. The approximate solutions of these equations
using the method based on the extraction of leading singularities of the
amplitude are obtained. The paper is devoted to the calculation results
for the baryonia mass spectrum (Table \ref{tab1}). In conclusion, the
status of the considered model is discussed.

The model in question has only two parameters of previous paper \cite{4}:
gluon coupling constant $g_0=0.314$ and cutoff parameter $\Lambda_q=11$.
We used the cutoff $\Lambda_{qc}=5.18$ which is determined by
$M=4100\, MeV$ (the threshold is equal to $4130\, MeV$).

The quark masses of the model are $m_q=495\, MeV$ and $m_c=1655\, MeV$.
The estimation of theoretical error on the $S$-wave hexaquarks masses
is $1\, MeV$. This results was obtained by the choice of model parameters.

We consider 9 baryonia with the content $qqQ\bar q\bar q\bar q$ and the
spin-parities $J^P=0^-$, $1^-$, $2^-$. The isospins are equal to $\frac{1}{2}$,
$\frac{3}{2}$, $\frac{5}{2}$ (Table \ref{tab1}). We have predicted the
masses of baryonia containing $c$ quark using the coupled-channel formalism.
We believe that the prediction for the $S$-wave charmed baryonia is based on
the relativistic kinematics and dynamics which allow us to take into account
the relativistic corrections.

\begin{table}
\caption{$qqQ\bar q\bar q\bar q$, $q=u, d$, $Q=c$. Parameters of model: cutoff
$\Lambda=11.0$, $\Lambda_{qc}=5.18$, gluon coupling constant $g=0.314$.
Quark masses $m_q=495\, MeV$, $m_c=1655\, MeV$.}
\label{tab1}
\begin{tabular}{|c|c|c|c|c|}
\hline
Quark content & $I$ & $J$ & Baryonium & Mass (MeV) \\
\hline
$uuc\,\, \bar u\bar u\bar u$, $ddc\,\, \bar d\bar d\bar d$, & $\frac{1}{2}$; $\frac{5}{2}$ & 0 &
$\Sigma^*_c \bar \Delta$, $\Delta \bar \Sigma^*_c$ & 3305 \\
$uuu\,\, \bar u\bar u\bar c$, $ddd\,\, \bar d\bar d\bar c$; &  & 1 &
$\Sigma_c \bar \Delta$, $\Delta \bar \Sigma_c$, $\Sigma^*_c \bar \Delta$, $\Delta \bar \Sigma^*_c$ & 3293 \\
$uuc\,\, \bar d\bar d\bar d$, $ddc\,\, \bar u\bar u\bar u$, & & 2 &
$\Sigma_c \bar \Delta$, $\Delta \bar \Sigma_c$, $\Sigma^*_c \bar \Delta$, $\Delta \bar \Sigma^*_c$ & 3303 \\
$ddd\,\, \bar u\bar u\bar c$, $uuu\,\, \bar d\bar d\bar c$ & & & & \\
\hline
$uuc\,\, \bar u\bar u\bar d$, $ddc\,\, \bar u\bar d\bar d$, & $\frac{1}{2}$; $\frac{3}{2}$ & 0 &
$\Sigma_c \bar N$, $N \bar \Sigma_c$, $\Sigma^*_c \bar \Delta$, $\Delta \bar \Sigma^*_c$ & 3317 \\
$uud\,\, \bar u\bar u\bar c$, $udd\,\, \bar d\bar d\bar c$; &  & 1 &
$\Sigma_c \bar N$, $N \bar \Sigma_c$, $\Sigma_c \bar \Delta$, $\Delta \bar \Sigma_c$,
$\Sigma^*_c \bar N$, $N \bar \Sigma^*_c$, $\Sigma^*_c \bar \Delta$, $\Delta \bar \Sigma^*_c$ & 3316 \\
$uuc\,\, \bar u\bar d\bar d$, $ddc\,\, \bar u\bar u\bar d$, & & 2 &
$\Sigma_c \bar \Delta$, $\Delta \bar \Sigma_c$,
$\Sigma^*_c \bar N$, $N \bar \Sigma^*_c$, $\Sigma^*_c \bar \Delta$, $\Delta \bar \Sigma^*_c$ & 3329 \\
$udd\,\, \bar u\bar u\bar c$, $uud\,\, \bar d\bar d\bar c$ & & & & \\
\hline
$udc\,\, \bar u\bar u\bar u$, $udc\,\, \bar d\bar d\bar d$, & $\frac{3}{2}$ & 0 &
$\Sigma^*_c \bar \Delta$, $\Delta \bar \Sigma^*_c$ & 3338 \\
$uuu\,\, \bar u\bar d\bar c$, $ddd\,\, \bar u\bar d\bar c$ & & 1, 2 &
$\Sigma_c \bar \Delta$, $\Delta \bar \Sigma_c$, $\Sigma^*_c \bar \Delta$, $\Delta \bar \Sigma^*_c$,
$\Lambda_c \bar \Delta$, $\Delta \bar \Lambda_c$ & 3309 \\
\hline
$udc\,\, \bar u\bar u\bar d$, $udc\,\, \bar u\bar d\bar d$, & $\frac{1}{2}$ & 0 &
$\Sigma_c \bar N$, $N \bar \Sigma_c$, $\Lambda_c \bar N$, $N \bar \Lambda_c$,
$\Sigma^*_c \bar \Delta$, $\Delta \bar \Sigma^*_c$ & 3331 \\
$uud\,\, \bar u\bar d\bar c$, $udd\,\, \bar u\bar d\bar c$ & & 1&
$\Sigma_c \bar N$, $N \bar \Sigma_c$, $\Sigma_c \bar \Delta$, $\Delta \bar \Sigma_c$,
$\Sigma^*_c \bar N$, $N \bar \Sigma^*_c$, & \\
& & &  $\Sigma^*_c \bar \Delta$, $\Delta \bar \Sigma^*_c$,
$\Lambda_c \bar N$, $N \bar \Lambda_c$, $\Lambda_c \bar \Delta$, $\Delta \bar \Lambda_c$ & 3331 \\
& & 2 & $\Sigma_c \bar \Delta$, $\Delta \bar \Sigma_c$,
$\Sigma^*_c \bar N$, $N \bar \Sigma^*_c$, $\Sigma^*_c \bar \Delta$, $\Delta \bar \Sigma^*_c$,
$\Lambda_c \bar \Delta$, $\Delta \bar \Lambda_c$ & 3361 \\
\hline
\end{tabular}
\end{table}

The quark pairs $Qq$ use there not so many, therefore the baryonium masses
cannot increase enough with the decreasing of the cutoff $\Lambda_{qc}$.

The degeneration of baryonium masses with the different spin-parities
$J^P=0^-$, $1^-$ was obtained. We cannot also calculate the bound states
of baryonia with $J^P=3^-$.

The baryonium state $\Sigma_c \bar \Delta$ $(uuc \, \bar d \bar d \bar d)$
for the spin-parities $J^P=0^-$, $1^-$, $2^-$ is calculated with the nine
subamplitudes: seven $\alpha_1$ (similar to $\alpha_1^{1^{uu}}$) and
two $\alpha_2^{1^{uu}1^{\bar d \bar d}}$, $\alpha_2^{0^{uc}1^{\bar d \bar d}}$.

The baryonium $\Sigma_c \bar \Delta$ $(uuc \, \bar u \bar d \bar d)$
consists of 16 subamplitudes with the spin-parities $J^P=0^-$, $1^-$;
12 $\alpha_1$ and 4 $\alpha_2$: $\alpha_2^{1^{uu}1^{\bar d \bar d}}$,
$\alpha_2^{1^{uu}0^{\bar u \bar d}}$, $\alpha_2^{0^{uc}1^{\bar d \bar d}}$,
$\alpha_2^{0^{uc}0^{\bar u \bar d}}$. For the case of the spin-parity
$J^P=2^-$ the subamplitude $\alpha_2^{0^{uc}0^{\bar u \bar d}}$ is
absent. The states with spin-parities $J^P=0^-$, $1^-$, $2^-$
$(udc \, \bar u \bar u \bar u)$ are constructed with 13 subamplitudes:
10 $\alpha_1$ and 3 $\alpha_2$: $\alpha_2^{0^{ud}1^{\bar u \bar u}}$,
$\alpha_2^{0^{uc}1^{\bar u \bar u}}$, $\alpha_2^{0^{dc}1^{\bar u \bar u}}$.
The baryonium $udc \, \bar u \bar u \bar d$ for the spin-parities
$J^P=0^-$, $1^-$ takes into account 23 subamplitudes: 17 $\alpha_1$ and
6 $\alpha_2$; $\alpha_2^{0^{ud}0^{\bar u \bar d}}$,
$\alpha_2^{0^{uc}0^{\bar u \bar d}}$, $\alpha_2^{0^{dc}0^{\bar u \bar d}}$,
$\alpha_2^{0^{ud}1^{\bar u \bar u}}$, $\alpha_2^{0^{uc}1^{\bar u \bar u}}$,
$\alpha_2^{0^{dc}1^{\bar u \bar u}}$. For the case $J^P=2^-$ the
subamplitudes $\alpha_2^{0^{ud}0^{\bar u \bar d}}$,
$\alpha_2^{0^{uc}0^{\bar u \bar d}}$, $\alpha_2^{0^{dc}0^{\bar u \bar d}}$
are absent.

The system of equations of the baryonium $\Sigma_c \bar \Delta$
$(uuc \, \bar d \bar d \bar d)$ for the spin-parity $J^P=1^-$
(as the example) was constructed:

\begin{eqnarray}
%1
\label{1}
\alpha_1^{1^{uu}}&=&\lambda+2\, \alpha_1^{0^{uc}} I_1(1^{uu}0^{uc})
+6\, \alpha_1^{1^{u\bar d}} I_1(1^{uu}1^{u\bar d})
+6\, \alpha_1^{0^{u\bar d}} I_1(1^{uu}0^{u\bar d})\, ,
\\
&&\nonumber\\
%2
\label{2}
\alpha_1^{0^{uc}}&=&\lambda+\alpha_1^{1^{uu}} I_1(0^{uc}1^{uu})
+\alpha_1^{0^{uc}} I_1(0^{uc}0^{uc})
+3\, \alpha_1^{1^{u\bar d}} I_1(0^{uc}1^{u\bar d})
+3\, \alpha_1^{0^{u\bar d}} I_1(0^{uc}0^{u\bar d})
\nonumber\\
&&\nonumber\\
&+&3\, \alpha_1^{1^{c\bar d}} I_1(0^{uc}1^{c\bar d})
+3\, \alpha_1^{0^{c\bar d}} I_1(0^{uc}0^{c\bar d})\, ,
\\
&&\nonumber\\
%3
\label{3}
\alpha_1^{1^{\bar d\bar d}}&=&\lambda+2\, \alpha_1^{1^{\bar d\bar d}} I_1(1^{\bar d\bar d}1^{\bar d\bar d})
+4\, \alpha_1^{1^{u\bar d}} I_1(1^{\bar d\bar d}1^{u\bar d})
+4\, \alpha_1^{0^{u\bar d}} I_1(1^{\bar d\bar d}0^{u\bar d})
+2\, \alpha_1^{1^{c\bar d}} I_1(1^{\bar d\bar d}1^{c\bar d})
\nonumber\\
&&\nonumber\\
&+&2\, \alpha_1^{0^{c\bar d}} I_1(1^{\bar d\bar d}0^{c\bar d})\, ,
\\
&&\nonumber\\
%4
\label{4}
\alpha_1^{1^{u\bar d}}&=&\lambda+\alpha_1^{1^{uu}} I_1(1^{u\bar d}1^{uu})+\alpha_1^{0^{uc}} I_1(1^{u\bar d}0^{uc})
+2\, \alpha_1^{1^{\bar d\bar d}} I_1(1^{u\bar d}1^{\bar d\bar d})
+3\, \alpha_1^{1^{u\bar d}} I_1(1^{u\bar d}1^{u\bar d})
\nonumber\\
&&\nonumber\\
&+&3\, \alpha_1^{0^{u\bar d}} I_1(1^{u\bar d}0^{u\bar d})
+\alpha_1^{1^{c\bar d}} I_1(1^{u\bar d}1^{c\bar d})+\alpha_1^{0^{c\bar d}} I_1(1^{u\bar d}0^{c\bar d})
+2\, \alpha_2^{1^{uu}1^{\bar d\bar d}} I_2(1^{u\bar d}1^{uu}1^{\bar d\bar d})
\nonumber\\
&&\nonumber\\
&+&2\, \alpha_2^{0^{uc}1^{\bar d\bar d}} I_2(1^{u\bar d}0^{uc}1^{\bar d\bar d})\, ,
\\
&&\nonumber\\
%5
\label{5}
\alpha_1^{0^{u\bar d}}&=&\lambda+\alpha_1^{1^{uu}} I_1(0^{u\bar d}1^{uu})+\alpha_1^{0^{uc}} I_1(0^{u\bar d}0^{uc})
+2\, \alpha_1^{1^{\bar d\bar d}} I_1(0^{u\bar d}1^{\bar d\bar d})
+3\, \alpha_1^{1^{u\bar d}} I_1(0^{u\bar d}1^{u\bar d})
\nonumber\\
&&\nonumber\\
&+&3\, \alpha_1^{0^{u\bar d}} I_1(0^{u\bar d}0^{u\bar d})
+\alpha_1^{1^{c\bar d}} I_1(0^{u\bar d}1^{c\bar d})+\alpha_1^{0^{c\bar d}} I_1(0^{u\bar d}0^{c\bar d})
+2\, \alpha_2^{1^{uu}1^{\bar d\bar d}} I_2(0^{u\bar d}1^{uu}1^{\bar d\bar d})
\nonumber\\
&&\nonumber\\
&+&2\, \alpha_2^{0^{uc}1^{\bar d\bar d}} I_2(0^{u\bar d}0^{uc}1^{\bar d\bar d})\, ,
\\
&&\nonumber\\
%6
\label{6}
\alpha_1^{1^{c\bar d}}&=&\lambda+2\, \alpha_1^{1^{\bar d\bar d}} I_1(1^{c\bar d}1^{\bar d\bar d})
+2\, \alpha_1^{1^{u\bar d}} I_1(1^{c\bar d}1^{u\bar d})
+2\, \alpha_1^{0^{u\bar d}} I_1(1^{c\bar d}0^{u\bar d})
+2\, \alpha_1^{1^{c\bar d}} I_1(1^{c\bar d}1^{c\bar d})
\nonumber\\
&&\nonumber\\
&+&2\, \alpha_1^{0^{c\bar d}} I_1(1^{c\bar d}0^{c\bar d})
+4\, \alpha_2^{0^{uc}1^{\bar d\bar d}} I_2(1^{c\bar d}0^{uc}1^{\bar d\bar d})\, ,
\\
&&\nonumber\\
%7
\label{7}
\alpha_1^{0^{c\bar d}}&=&\lambda+2\, \alpha_1^{1^{\bar d\bar d}} I_1(0^{c\bar d}1^{\bar d\bar d})
+2\, \alpha_1^{1^{u\bar d}} I_1(0^{c\bar d}1^{u\bar d})
+2\, \alpha_1^{0^{u\bar d}} I_1(0^{c\bar d}0^{u\bar d})
+2\, \alpha_1^{1^{c\bar d}} I_1(0^{c\bar d}1^{c\bar d})
\nonumber\\
&&\nonumber\\
&+&2\, \alpha_1^{0^{c\bar d}} I_1(0^{c\bar d}0^{c\bar d})
+4\, \alpha_2^{0^{uc}1^{\bar d\bar d}} I_2(0^{c\bar d}0^{uc}1^{\bar d\bar d})\, ,
\\
&&\nonumber\\
%8
\label{8}
\alpha_2^{1^{uu}1^{\bar d\bar d}}&=&\lambda+2\, \alpha_1^{0^{uc}} I_4(1^{uu}1^{\bar d\bar d}0^{uc})
+2\, \alpha_1^{1^{\bar d\bar d}} I_4(1^{uu}1^{\bar d\bar d}1^{\bar d\bar d})
+4\, \alpha_1^{1^{u\bar d}} I_3(1^{uu}1^{\bar d\bar d}1^{u\bar d})
\nonumber\\
&&\nonumber\\
&+&4\, \alpha_1^{0^{u\bar d}} I_3(1^{uu}1^{\bar d\bar d}0^{u\bar d})
+4\, \alpha_2^{0^{uc}1^{\bar d\bar d}} I_6(1^{uu}1^{\bar d\bar d}0^{uc}1^{\bar d\bar d})\, ,
\\
&&\nonumber\\
%9
\label{9}
\alpha_2^{0^{uc}1^{\bar d\bar d}}&=&\lambda+\alpha_1^{1^{uu}} I_4(0^{uc}1^{\bar d\bar d}1^{uu})
+\alpha_1^{0^{uc}} I_4(0^{uc}1^{\bar d\bar d}0^{uc})
+2\, \alpha_1^{1^{\bar d\bar d}} I_4(1^{\bar d\bar d}0^{uc}1^{\bar d\bar d})
\nonumber\\
&&\nonumber\\
&+&2\, \alpha_1^{1^{u\bar d}} I_3(0^{uc}1^{\bar d\bar d}1^{u\bar d})
+2\, \alpha_1^{0^{u\bar d}} I_3(0^{uc}1^{\bar d\bar d}0^{u\bar d})
+2\, \alpha_1^{1^{c\bar d}} I_3(0^{uc}1^{\bar d\bar d}1^{c\bar d})
\nonumber\\
&&\nonumber\\
&+&2\, \alpha_1^{0^{c\bar d}} I_3(0^{uc}1^{\bar d\bar d}0^{c\bar d})
+2\, \alpha_2^{1^{uu}1^{\bar d\bar d}} I_6(0^{uc}1^{\bar d\bar d}1^{uu}1^{\bar d\bar d})
+2\, \alpha_2^{0^{uc}1^{\bar d\bar d}} I_6(0^{uc}1^{\bar d\bar d}0^{uc}1^{\bar d\bar d})\, .
\end{eqnarray}

We used the functions $I_1$, $I_2$, $I_3$, $I_4$, $I_6$ similar to the paper \cite{3}.

The poles of the reduced amplitudes $\alpha_l$ correspond to the bound states
and determine the masses of the charmed baryonia.

In Table \ref{tab1} the calculated masses of nonstrange baryonia with
the open charm are shown.

We predict the mass of lowest charmed baryonium with the isospin
$I=\frac{1}{2}$ and the spin-parity $J^P=1^-$ ($M=3293\, MeV$).

The known way with which to calculate the low-energy properties of hadronic
systems rigorously is Lattice QCD (LQCD) \cite{5, 6}. In LQCD calculations,
the quark and gluon fields are defined on a discretized space-time of finite
volume of the lattice volume, such deviation can be systematically removed
by reducing the lattice spacing, increasing the lattice volume and
extrapolating to the continuum and infinite volume limits using the known
dependences determined with effective field theory (EFT) \cite{7, 8, 9}.

We try to consider the tasks which are similar to the Lattice calculations.

\begin{acknowledgments}
The work was carried out with the support of the Russian Ministry of Education
(grant 2.1.1.68.26).
\end{acknowledgments}

\end{document}